\documentstyle[11pt,amssymb,epsf]{article}

\def\crampest{\medmuskip = 1mu plus 1mu minus 1mu}

\def\ben{\begin{equation}}
\def\een{\end{equation}}

  \let\n=\nu

\let\C=\Chi

\def\nn{\nonumber} \def\bd{\begin{document}} \def\ed{\end{document}}
\def\ds{\documentstyle} \let\fr=\frac \let\bl=\bigl \let\br=\bigr
\let\Br=\Bigr \let\Bl=\Bigl
\let\bm=\bibitem
\let\na=\nabla
\let\pa=\partial \let\ov=\overline
\newcommand{\be}{\begin{equation}}
\newcommand{\ee}{\end{equation}}
\def\ba{\begin{array}}
\def\ea{\end{array}}
\def\ft#1#2{{\textstyle{{\scriptstyle #1}\over {\scriptstyle #2}}}}
\def\fft#1#2{{#1 \over #2}}
\def\del{\partial}
\def\vp{\varphi}
\def\sst#1{{\scriptscriptstyle #1}}
\def\oneone{\rlap 1\mkern4mu{\rm l}}
\def\td{\tilde}
\def\wtd{\widetilde}
\def\ie{\rm i.e.\ }
\def\dalemb#1#2{{\vbox{\hrule height .#2pt
        \hbox{\vrule width.#2pt height#1pt \kern#1pt
                \vrule width.#2pt}
        \hrule height.#2pt}}}
\def\square{\mathord{\dalemb{6.8}{7}\hbox{\hskip1pt}}}
\newcommand{\ho}[1]{$\, ^{#1}$}
\newcommand{\hoch}[1]{$\, ^{#1}$}
\newcommand{\bea}{\begin{eqnarray}}
\newcommand{\eea}{\end{eqnarray}}
\newcommand{\ra}{\rightarrow}
\newcommand{\lra}{\longrightarrow}
\newcommand{\Lra}{\Leftrightarrow}
\newcommand{\bp}{\tilde \beta^\prime}
\newcommand{\tr}{{\rm tr} }
\newcommand{\Tr}{{\rm Tr} }
\def\0{{\sst{(0)}}}
\def\1{{\sst{(1)}}}
\def\2{{\sst{(2)}}}
\def\3{{\sst{(3)}}}
\def\4{{\sst{(4)}}}
\def\5{{\sst{(5)}}}
\def\6{{\sst{(6)}}}
\def\7{{\sst{(7)}}}
\def\8{{\sst{(8)}}}
\def\n{{\sst{(n)}}}
\def\cA{{{\cal A}}}
\def\cB{{{\cal B}}}
\def\cF{{{\cal F}}}
\def\cH{{{\cal H}}}
\def\tV{\widetilde V}
\def\tW{\widetilde W}
\def\tH{\widetilde H}
\def\tE{\widetilde E}
\def\tF{\widetilde F}
\def\tA{\widetilde A}
\def\im{{i}}
\def\tY{{{\wtd Y}}}
\def\ep{{\epsilon}}
\def\vep{{\varepsilon}}
\def\R{\rlap{\rm I}\mkern3mu{\rm R}}
\def\bD{{{\bar D}}}

\def\R{\rlap{\rm I}\mkern3mu{\rm R}}
\def\bD{{{\bar D}}}
\def\R{{{\Bbb R}}}
\def\C{{{\Bbb C}}}
\def\H{{{\Bbb H}}}
\def\CP{{{\Bbb C}{\Bbb P}}}
\def\RP{{{\Bbb R}{\Bbb P}}}
\def\Z{{{\Bbb Z}}}
\def\bA{{{\Bbb A}}}
\def\bB{{{\Bbb B}}}
\def\bC{{{\Bbb C}}}
\def\bD{{{\Bbb D}}}
\def\bE{{{\Bbb E}}}
\def\bZ{{{\Bbb Z}}}
\def\Re{{{\frak{Re}}}}
\def\Im{{{\frak{Im}}}}
\def\cosec{{\,\hbox{cosec}\,}}
\def\Gm{{\Gamma_{\!\! -}}}
\def\Gp{{\Gamma_{\!\! +}}}
\def\stan{{standard }}
\def\nonstan{{supernumerary }}

\newcommand{\tamphys}{\it Center for Theoretical Physics,
Texas A\&M University, College Station, TX 77843}

\newcommand{\upenn}{\it Department of Physics and Astronomy,\\ University
of Pennsylvania, Philadelphia, PA 19104}

\newcommand{\brussels}{\it Physique Th\'eorique et Math\'ematique,
Universit\'e Libre de Bruxelles,\\ Campus Plaine C.P. 231, B-1050
Bruxelles, Belgium}

\newcommand{\auth}{Z.-W. Chong\hoch{\ddagger1}, M. Cveti\v c\hoch{*2},  
H. L\"u\hoch{\ddagger1} and C.N. Pope\hoch{\ddagger1}}

\begin{document}

\begin{flushright}
MIFP-04-24\ \ \ UPR-1100-T \\
{\bf hep-th/0412094}\\
December\  2004
\end{flushright}

\vspace{10pt}

\begin{center}

{\large {\bf Non-extremal Charged Rotating Black Holes in Seven-Dimensional
Gauged Supergravity}}

\vspace{20pt}
\auth

\vspace{10pt}{\hoch{\ddagger}\it George P. \& Cynthia W. Mitchell
Institute for Fundamental Physics,\\ Texas A\& M University,
College Station, TX 77843-4242, USA}

\vspace{10pt}{\hoch{*}\it Department of Physics and Astronomy,\\
University of Pennsylvania, Philadelphia, PA 19104, USA}


%
%
%

\vspace{20pt}


\begin{abstract}

    We obtain the solution for non-extremal charged rotating black
holes in seven-dimensional gauged supergravity, in the case where the
three rotation parameters are set equal.  There are two independent
charges, corresponding to gauge fields in the $U(1)\times U(1)$
abelian subgroup of the $SO(5)$ gauge group.  A new feature in these
solutions, not seen previously in lower-dimensional examples, is that
the first-order ``odd-dimensional self-duality'' equation for the
4-form field strength plays a non-trivial r\^ole.  We also study the
BPS limit of our solutions where the black holes become
supersymmetric.  Our results are of significance for the
AdS$_7$/CFT$_6$ correspondence in M-theory.

\end{abstract}
\end{center}

{\vfill\leftline{}\vfill \vskip 10pt \footnoterule {\footnotesize
{\footnotesize
\hoch{1} Research supported in part by DOE grant
DE-FG03-95ER40917.}\vskip 2pt
\hoch{2} Research supported in part by DOE grant
DE-FG02-95ER40893, NSF grant INTO3-24081, and the\\
$\phantom{xxxxi}$  Fay R. and Eugene L.
Langberg Chair.}\vskip 2pt
}

\pagebreak

\newpage

\section{Introduction}

   Charged black holes in gauged supergravities provide gravitational
backgrounds of importance in the study of the AdS/CFT correspondence.
Non-extremal black hole solutions are relevant for studying the
dual field theory at non-zero temperature.  This has been discussed
extensively for static AdS black holes in, for example,
\cite{Sabraetal,CGI,CGII}.  See also \cite{Buchel,SabraII,gubheck},
for recent related work.  For non-extremal charged rotating black
holes in gauged supergravities, little has been known until recently.
In \cite{cvlupo1,cvlupo2} the first examples of non-extremal rotating
charged AdS black holes in five-dimensional ${\cal N}=4$ gauged
supergravity were obtained, in the special case where the two angular
momenta $J_i$ are set equal.  These solutions are characterised by
their mass, three electromagnetic charges, and the angular momentum
parameter $J=J_1=J_2$.  By taking appropriate limits, one obtains the
various supersymmetric charged rotating $D=5$ black holes obtained in
\cite{gutreal1,gutreal2,klesab}.  If instead the charges are set to
zero, the solutions reduce to the rotating AdS$_5$ black hole
constructed in \cite{hhtr}, with $J_1=J_2$.  In four dimensions, the
charged Kerr-Newman-AdS black hole solution of the Einstein-Maxwell
system with a cosmological constant has long been known
\cite{carter1,carter2}.  This can be viewed as a solution in gauged
${\cal N}=8$ supergravity, in which the four electromagnetic fields in
the $U(1)^4$ abelian subgroup of the $SO(8)$ gauge group are set
equal.  Recently, a more general class of non-extremal charged rotating
solutions in the four-dimensional gauged theory were constructed, in 
which the four electric charges are set pairwise equal \cite{chcvlupo}.  

   Another case of interest from the AdS/CFT perspective is
non-extremal charged rotating black holes in seven-dimensional gauged
supergravity, and this forms the subject of the present paper.  The
maximally-supersymmetric theory has ${\cal N}=4$ supersymmetry, and
the gauge group is $SO(5)$ \cite{d7sugra}.  It was shown in
\cite{vann1,vann2} that this theory can be obtained as a consistent
reduction of eleven-dimensional supergravity on $S^4$.  A convenient
presentation of the Lagrangian for the bosonic sector, in the
conventions we shall be using, appears in \cite{cvlupotr}.  The theory
is capable of supporting black holes carrying two independent electric
charges, carried by gauge fields in the $U(1)\times U(1)$ abelian
subgroup of the full $SO(5)$ gauge group.  For the purposes of
discussing the solutions, it therefore suffices to perform a
(consistent) truncation of the full supergravity theory to the
relevant sector, in which all except the $U(1)\times U(1)$ subgroup of
gauge fields are set to zero.  The fields retained in the consistent
truncation comprise the metric, two dilatons, the $U(1)\times U(1)$
gauge fields and a 4-form field strength that satisfies an
odd-dimensional self-duality equation.  The equations of motion can be
derived from the Lagrangian
\bea
{\cal L}_7 &=& R\, {*\oneone} - \ft12 {*d\varphi_i}\wedge d\varphi_i -
  \ft12 \sum_{i=1}^2 X_i^{-2}\, {*F_\2^i}\wedge F_\2^i 
    -\ft12 (X_1\, X_2)^2\, {*F_\4}\wedge F_\4\nn\\
&& + 2g^2\, [(X_1\, X_2)^{-4} -8 X_1\, X_2 -4 X_1^{-1} X_2^{-2} - 
              4 X_1^{-2}\, X_2^{-1}] \nn\\
&&-g\, F_\4\wedge A_\3 
   + F_\2^1\wedge F_\2^2\wedge A_\3\,,\label{d7lag}
\eea
where 
\bea
F_\2^i &=& dA_\1^i\,,\qquad F_\4=dA_\3\,,\nn\\
X_1 &=& e^{-\ft1{\sqrt2}\, \varphi_1 -\ft1{\sqrt{10}}\, \varphi_2}
\,,\qquad
X_2 = e^{\ft1{\sqrt2}\, \varphi_1 -\ft1{\sqrt{10}}\, \varphi_2}\,,
\eea
together with a first-order ``odd-dimensional self-duality'' equation
to be imposed after the variation of the Lagrangian.  This condition
is most conveniently stated by introducing an additional 2-form
potential $A_\2$, which can be gauged away in the gauged theory, and
defining
\be
F_\3 = dA_\2 - \ft12 A_\1^1\wedge dA_\1^2 - \ft12 A_\1^2\wedge dA_\1^1\,.
\ee
The odd-dimensional self-duality equation then reads\footnote{Note that
if $g\ne 0$, one can absorb $A_2$ by making a gauge transformation of
$A_\3$.  If, on the other hand, $g=0$, then (\ref{odddim}) just becomes
the defining equation for $F_\3$ as the dual of $F_\4$.  When $g=0$
one can equivalently work either with $A_\3$, or with $A_\2$ in a dual 
formulation of the theory.}
\be
(X_1\, X_2)^2\, {* F_\4} = -2g\, A_\3 - F_\3\,.\label{odddim}
\ee
This is a first integral of the equation of motion for
$A_\3$ that follows directly from (\ref{d7lag}).  (Note that one can
alternatively write a Lagrangian that yields the equations of motion
directly, with no need for an additional constraint.  See, for example,
\cite{d7sugra,cvlupotr}.) 

   The fact that the 3-form $A_\3$ satisfies the odd-dimensional
self-duality equation presents an interesting new challenge when
constructing the charged rotating solutions in the gauged
seven-dimensional supergravity.  The trickiest part of finding charged
rotating solutions in any of the gauged supergravities is that one has
little {\it a priori} guidance as to how the dimensionless quantity
$a\, g$ enters the solution, where $a$ is the rotation parameter and
$g$ the gauge coupling constant.  In the cases that have been
constructed previously, in five dimensions \cite{cvlupo1,cvlupo2} and
in four dimensions \cite{chcvlupo}, the gauge coupling constant appeared
always quadratically in the relevant equations of motion, and thus the
dimensionless product entered the solutions in the combination $a^2\,
g^2$.  In seven dimensions, by contrast, the gauge coupling constant
$g$ appears linearly in the odd-dimensional self-duality equation
(\ref{odddim}), and so in turn the solution involves linear powers of
the product $a\, g$.  This considerably complicates the task of
parameterising possible forms for the solution, in the process of
formulating a conjecture and then verifying that it works.  It is
intriguing that having found the charged rotating black hole, we have
the rather uncommon situation of obtaining a solution of seven-dimensional 
gauged supergravity in which the 
odd-dimensional self-duality equation (\ref{odddim}) is satisfied
in a non-trivial way.

    In the following sections, we shall construct non-extremal charged
rotating black-hole solutions in the seven-dimensional gauged
supergravity.  In our approach, we begin with the previously-known
charged rotating solutions in the ungauged theory.  Then, we formulate
a conjecture for the generalisation to the gauged theory, and verify
it by an explicit checking of all the equations of motion.  The
charged solutions in the ungauged supergravity were constructed (with
the full complement of three independent rotation parameters) in
\cite{cvetyoum}.  In order to give a uniform presentation of our
results, and also to eliminate some typographical errors that arose in
\cite{cvetyoum}, we begin in section 2 by rederiving the charged
rotating black holes in the ungauged seven-dimensional supergravity,
in the special case we are addressing in this paper where the three
angular momenta are set equal.  Then, in section 3, we formulate our
conjectured generalisation to the gauged supergravity theory, and
verify that it does indeed satisfy the equations of motion.  As 
well as obtaining the non-extremal solutions with two independent
charges, we also present a somewhat simpler form of the metric in
the special case where the two charges are set equal.  In section 4,
we discuss the BPS limit, showing how supersymmetric rotating black
hole solutions in seven-dimensional gauged supergravity arise for
a suitable restriction of the parameters.   The paper ends with
conclusions in section 5.

\section{Charged Rotating Black Holes in the Ungauged Theory}

    Charged solutions in ungauged supergravity can be obtained from
uncharged ones by making use of global symmetries of the theory,
employed as solution-generating transformations.  In the present case,
one way of doing this is to recognise that in the ungauged ($g=0$)
limit, the seven-dimensional theory described by (\ref{d7lag}) can be
obtained as the dimensional reduction of the eight-dimensional
``bosonic string'' theory described by
\be
{\cal L}_8 = \hat R\, {\hat * \oneone} - \ft12 {\hat *d\varphi}\wedge 
          d\varphi - \ft12 e^{-\ft2{\sqrt3} \varphi}\, 
{\hat * \hat F_\3}\wedge 
              \hat F_\3\,.
\ee
This yields the seven-dimensional theory in a formulation in which the 4-form 
$F_\4$ has been dualised to the 3-form $F_\3$ (see footnote 1).  The
strategy for introducing charges is then to begin with an uncharged, 
Ricci-flat solution in seven dimensions, take its product with a circle,
and hence obtain a Ricci-flat solution of the eight-dimensional theory.
Next, one performs a Lorentz transformation in the $(t,z)$ plane, 
with Lorentz boost parameter $\delta_1$, where
\be
t\longrightarrow t\, \cosh\delta_1 + z\, \sinh\delta_1\,,\qquad
z\longrightarrow z\, \cosh\delta_1 + t\, \sinh\delta_1\,,
\ee
where $z$ is the circle coordinate of the eighth dimension.  Upon
reduction to $D=7$ on the Lorentz-transformed circle coordinate $z$,
one obtains a seven-dimensional solution in which the Kaluza-Klein
vector carries an electric charge.  The next step is to use the
discrete $Z_2$ subgroup of the seven-dimensional global symmetry group
that exchanges the Kaluza-Klein and winding vectors.  This allows one to
repeat the lifting, Lorentz boosting and reduction steps, with a 
second boost parameter $\delta_2$, thereby ending up with a seven-dimensional
solution where each of the Kaluza-Klein and winding vectors carries an
electric charge.

   In principle we can apply this charge-generating procedure starting 
from any Ricci-flat metric in seven dimensions.  In our present case, we take
as our starting point the generalisation of the rotating Kerr black hole
to seven dimensions, obtained by Myers and Perry \cite{myeper}.  The most
general such solution has three independent rotation parameters in the
three orthogonal 2-planes of its six-dimensional transverse space.  For
reasons of simplicity, we restrict attention to the case where the three
rotation parameters are set equal.   

   The uncharged seven-dimensional rotating black hole, with the 
three rotation parameters set equal, can be written as
\be
ds_7^2 =  - dt^2 + \fft{2m}{\rho^4}\, (dt-a\, \sigma)^2  + 
  \fft{\rho^4\, dr^2}{V-2m}  + \rho^2\, (d\Sigma_2^2 + \sigma^2)\,,
\ee
where
\be
\rho^2\equiv (r^2+a^2)\,,\qquad V\equiv \fft1{r^2}\, (r^2+a^2)^3\,,
\ee
$d\Sigma_2^2$ is the standard Fubini-Study metric on $\CP^2$, and
$\sigma$ is the connection on the $U(1)$ fibre over $\CP^2$ whose total
bundle is the unit 5-sphere.  Thus we may write \cite{gibpop}
\bea
d\Sigma_2^2 &=& d\xi^2 + \ft14 \sin^2\xi\, (\sigma_1^2 + \sigma_2^2) +
   \ft14 \sin^2\xi\, \cos^2\xi\, \sigma_3^2\,,\nn\\
\sigma &=& d\tau + \ft12 \sin^2\xi\, \sigma_3\,,
\eea
where $\sigma_i$ denotes a set of left-invariant 1-forms on $SU(2)$, 
satisfying $d\sigma_i= -\ft12 \ep_{ijk}\, \sigma_j\wedge \sigma_k$.  Note
that we have
\be
d\sigma = 2 J\,,
\ee
where $J$ is the K\"ahler form on $\CP^2$.  

   After implementing the sequence of steps described above in order
to introduce electric charges, we find that the charged
rotating non-extremal seven-dimensional black holes are given by
\bea
ds_7^2 &=& (H_1\, H_2)^{1/5}\, 
   \Big[ -\fft{\rho^4-2m}{\rho^4\, H_1\, H_2}\, dt^2 - 
   \fft{4m\, a}{\rho^4\, H_1\, H_2}\, dt\, \sigma + 
   \fft{2m\, a^2}{\rho^4\, H_1\, H_2}\,
\Big(1- \fft{2m\, s_1^2\, s_2^2}{\rho^4}\Big)
  \,\sigma^2\nn\\
&& \qquad\qquad \quad + \rho^2\, (d\Sigma_2^2 + \sigma^2) +
\fft{\rho^4\, dr^2}{V-2m}\Big]\,,\nn\\
A_\1^1 &=& \fft{2m\, s_1}{\rho^4\, H_1}\, 
(c_1\, dt - a\, c_2\, \sigma)\,,\qquad
A_\1^2 = \fft{2m\, s_2}{\rho^4\, H_2}\, 
(c_2\, dt - a\, c_1\, \sigma)\,,\nn\\
A_\2 &=& \fft{m\, a\, s_1\, s_2}{\rho^4}\, \Big(\fft1{H_1} + \fft1{H_2}\Big) 
\, dt\wedge \sigma\,,\nn\\
X_i &=& (H_1\, H_2)^{2/5}\, H_i^{-1}\,,\label{ungaugedsol}
\eea
where 
\be
H_i= 1+ \fft{2m\, s_i^2}{\rho^4}\,,
\ee
and we have defined
\be
s_i \equiv \sinh\delta_i\,,\qquad c_i\equiv \cosh\delta_i\,.
\ee

   (A different solution-generating technique, making use of global
symmetries of the three-dimensional theory obtained by dimensional
reduction, was used in \cite{cvetyoum} to construct rotating charged
black holes in $D$-dimensional supergravities for $4\le D\le 9$, with
2 independent charges and $[(D-1)/2]$ independent angular momenta.
When the three angular momenta in the $D=7$ solution are set equal,
the situation considered in \cite{cvetyoum} reduces to the one we have
considered in this paper.\footnote{The general solution in \cite{cvetyoum}
(eq.~(12) of \cite{cvetyoum}) has a few typographical errors: a term
$2N\, \ell_i^2\, \mu_i^2$ in the metric coefficient for $d\phi_i^2$ should be
$2N\, \Delta\, \ell_i^2\, \mu_i^2$, the 2-form potential components 
$B_{\phi_i\, \phi_j}$ should be set to zero, and the quantity $m r$ 
in $B_{t\, \phi_i}$ should be $N$.})

   Note that the 3-form $F_\3=dA_\2 - \ft12 A_\1^1\wedge dA_\1^2 -
 \ft12 A_\1^2\wedge dA_\1^1$ can be dualised to the 4-form $F_\4=dA_\3$, 
in which case one has
\be
A_\3 = \fft{2m\, a\, s_1\, s_2}{r^2+a^2}\, \sigma\wedge J
\ee
in place of the expression for $A_\2$ in (\ref{ungaugedsol}).

\section{Charged Rotating Black Holes in the Gauged Theory}

   In this section, we construct non-extremal charged rotating
solutions in the gauged seven-dimensional supergravity theory.  Note
that the global symmetries that allowed us to generate charged
solutions from uncharged ones are broken in the gauged theory, and so
there is no longer a procedure available that delivers the charged
solutions by mechanical means.  Instead, we have constructed the
charged solutions by means of ``educated guesswork,'' followed by an
explicit verification that all the seven-dimensional equations of
motion are indeed satisfied.  In order to conjecture the form of the
solution, we have made extensive use of previously-known limiting cases,
including, especially, the charged solutions of the ungauged theory,
which we described in the previous section.  

   We find that the charged and rotating non-extremal black hole solution
of the seven-dimensional gauged supergravity is 
given by\footnote{It should be emphasised that in the solution
(\ref{2sol}), $A_\3$ is the potential for the fundamental field
$F_\4=dA_\3$ in the gauged supergravity, while, as discussed in
footnote 1, $A_\2$ is a term that could, if one wished, be viewed as
being absorbed into $A_\3$ via a gauge transformation of $A_\3$.  It
happens to be convenient to present it in the form we have done; we
are {\it not} saying that $A_\2$ is an independent fundamental field.}
\bea
ds_7^2 &=& (H_1 H_2)^{1/5}\, \Big[ -\fft{Y\, dt^2}{f_1\, \Xi_-^2} 
  + \fft{\rho^4\, d\rho^2}{Y}+ \fft{f_1}{\rho^4\, H_1 H_2 \, \Xi^2}\, 
   \Big(\sigma - \fft{2 f_2}{f_1}\, dt\Big)^2 + 
   \fft{\rho^2}{\Xi}\, d\Sigma_2^2\Big]\,,\nn\\
A_\1^i &=& \fft{2m\, s_i}{\rho^4\,\Xi\, H_i}\, 
(\alpha_i\, dt + \beta_i\, \sigma)
\,, \nn\\
A_\2 &=& \fft{m\, a\, s_1\, s_2}{\rho^4\, \Xi_-^2}\, 
    \Big(\fft1{H_1} + \fft1{H_2}\Big)\, dt\wedge \sigma\,,\qquad
 A_\3 = \fft{2m\, a\, s_1\, s_2}{\rho^2\, \Xi\, \Xi_-}\,
           \sigma\wedge J\,,\nn\\
X_i &=& (H_1 H_2)^{2/5}\, H_i^{-1}\,,\qquad 
H_i = 1 + \fft{2m\, s_i^2}{\rho^4}\,,\qquad \rho^2= r^2+a^2\,,\nn\\
\alpha_1&=& c_1 - \ft12 (1-\Xi_+^2)(c_1-c_2)\,,\qquad
\alpha_2= c_2 + \ft12 (1-\Xi_+^2)(c_1-c_2)\,,\nn\\
\beta_1 &=& -a\, \alpha_2\,,\qquad 
\beta_2 = -a\, \alpha_1\,,\nn\\
\Xi_\pm &=& 1 \pm a\, g\,,\qquad \Xi =  1 -a^2\, g^2
                         = \Xi_-\, \Xi_+\,,\label{2sol}
\eea
where the functions $f_1$, $f_2$ and $Y$ are given by
\crampest{
\bea
f_1 &=& \Xi\, \rho^6\, H_1 H_2 - 
     \fft{4\Xi_+^2\, m^2\, a^2\, s_1^2\, s_2^2}{\rho^4} +
 \ft12 m\, a^2\, \Big[4 \Xi_+^2 + 2 c_1\, c_2\,(1-\Xi_+^4) 
         + (1-\Xi_+^2)^2\, (c_1^2+c_2^2)\Big]\,,\nn\\
f_2 &=& -\ft12 g\, \Xi_+\, \rho^6\, H_1 H_2 + \ft14 m\, a\, \Big[
  2(1+\Xi_+^4)\, c_1\, c_2 + (1-\Xi_+^4)\, (c_1^2 + c_2^2)\Big]\,,\nn\\
Y &=& g^2\, \rho^8\, H_1 H_2 + \Xi\, \rho^6 
+ \ft12 m\, a^2\, \Big[ 4 \Xi_+^2 + 2(1-\Xi_+^4)\, c_1\, c_2 
+ (1-\Xi_+^2 )^2 (c_1^2 + c_2^2)\Big]\nn\\
&&-\ft12 m\, \rho^2 \,\Big[
  4\Xi + 2 a^2 g^2 (6+ 8 a g + 3 a^2 g^2)\, c_1\, c_2  -
  a^2 g^2 (2+a g)(2+3a g)(c_1^2+c_2^2)\Big]\,. 
\eea
}
It is a purely mechanical exercise, which we performed with the aid of
Mathematica, to verify that this configuration indeed satisfies the
equations of motion of seven-dimensional gauged supergravity,
following from (\ref{d7lag}) together with the odd-dimensional
self-duality equation (\ref{odddim}).  Note that as mentioned in the
introduction, unlike the charged rotating AdS black holes in $D=5$ and
$D=4$, the metric in $D=7$ depends on odd powers of $g$ as well as
even powers, in consequence of the odd-dimensional self-duality equation.

   If one specialises to the case where the two charges are set equal,
the solution may be written in a somewhat simpler form, as 
\bea
ds_7^2 &=& H^{2/5}\, \Big[ - \fft{V-2m}{\rho^4\, H^2\,\Xi^2}\,
(dt -a\, \sigma)^2
   + \fft1{r^2\, H^2\,\Xi^2}\, (h_1\, dt - h_2\, \sigma)^2 + 
    \fft{\rho^4\, dr^2}{V-2m} + \fft{\rho^2}{\Xi}\, d\Sigma_2^2\Big]
\,,\nn\\
A_\1^1 &=& A_\1^2 = \fft{2m\, s\, c}{\rho^4\,H\, \Xi}\,
(dt - a\, \sigma)\,,\nn\\
A_\2 &=& \fft{2 m\, s^2\, a}{\rho^4\, H\, \Xi_-^2}\, dt
\wedge\sigma\,,\nn\\
A_\3 &=& \fft{2m\, a\, s^2}{\Xi\, \Xi_-\, (r^2+a^2)}\, \sigma\wedge J\,,
\label{equalsol}
\eea
where
\bea
V &=& \fft1{r^2}\, \Big( (r^2+a^2)^3\, (1+g^2\, r^2) + 
  2g\, m\, (2g\, r^4 + 3 a^2\, g\, r^2 \
  - 2 a^3)\, s^2 + 4 g^2\, m^2\, s^4 \Big)\,,\nn\\
h_1 &=& -a + g\,r^2\, H + \fft{2a^2\, g\, m\, s^2}{\rho^4}\,,\qquad
h_2 = - a^2 - r^2\, H + \fft{2 a^3\, g\, m\, s^2}{\rho^4}\,,\nn\\
H&=&1 + \fft{2m s^2}{\rho^4}\,,\qquad\rho^2=r^2+a^2\,,\qquad
s\equiv \sinh\delta\,,\qquad c\equiv \cosh\delta\,,\qquad
\eea

\section{The Supersymmetric Limit}

   The charged rotating black hole solutions in seven-dimensional
gauged supergravity that we have derived in this paper are in general
non-extremal, with the mass and the electric charges freely
specifiable.  It is of interest also to study the extremal limit, in
which one obtains supersymmetric BPS black hole solutions.  For
simplicity, we shall just present the results for the case where the two
electric charges are set equal here.

   The criterion for supersymmetry is that there should exist
supersymmetry parameters $\epsilon$ such that the supersymmetry
variations of the spin-$\ft32$ and spin-$\ft12$ fields $\psi_\mu$ and
$\lambda_i$ in the gauged supergravity theory should vanish.  This is
most easily checked by looking at the integrability condition for the
spin-$\ft32$ field, and by looking directly at the transformation rule
for the spin-$\ft12$ field.  This latter, in the case where the
electric charges are set equal, takes the form
\be
\delta \lambda_i = -\ft14 \Gamma^\mu\,\ep\,  
       X^{-1}\, \del_\mu X + \ft{\im}{40}\, 
  X^{-1}\, F_{\mu\nu}\, \Gamma^{\mu\nu}\, \ep - \ft1{480}\, 
    X^2\, F_{\mu\nu\rho\sigma}\, \Gamma^{\mu\nu\rho\sigma}\, \ep 
+ \ft15 g\, (X-X^{-4})\, \ep\,.
\ee
By studying the eigenvalues of the matrix that acts on $\ep$, we find that
there can exist Killing spinors if the parameter $\delta$ satisfies
\be
\tanh\delta = \fft{\pm 1}{1+a\, g}\,,\label{tand}
\ee
which implies that 
\be
s\equiv \sinh\delta = \fft{\pm 1}{\sqrt{\Xi_+^2-1}}\,,\qquad
c\equiv \cosh\delta = \fft{\Xi_+}{\sqrt{\Xi_+^2-1}}\,.
\ee
Specifically, we find that the $8\times 8$ matrix has two zero eigenvalues
if equation (\ref{tand}) holds. 

   Our results for supersymmetric black holes reduce to previously-known
cases if we specialise to $g=0$ or $a=0$.  In either of these cases, we
find that the number of zero eigenvalues increases to four, with
the BPS condition (\ref{tand}) reducing now to the familiar one that
the ``extremality parameter'' is given by $\delta\rightarrow \pm\infty$
and $m\rightarrow 0$ with $m \sinh^2\delta$ fixed in the BPS
limit.  Thus we see that, as is the case also in four dimensions, 
BPS rotating black holes in gauged supergravity have only one half of
the supersymmetry that occurs if either the rotation or the gauge
coupling is set to zero.

       It is not uncommon, for certain ranges of the parameters, 
for a rotating black hole to have
naked closed timelike curves (CTCs).  In our solution, with the two
charges set equal, it is easy to see that
\bea
 H^{-2/5}\, g_{00} &=& \Big(\fft{4f_2^2}{\rho^4 H^2 \Xi^2} -
\fft{Y}{\Xi_{-}^2}\Big)\fft{1}{f_1} \nn\\
&=&\fft{2m (1- (\Xi_{+}^2-1)s^2) - \Xi_+^2 \rho^4}{\Xi^2
H^2 \rho^4}\,.
\eea
The horizon is located at the outer root of 
$Y=0$.  The absence of CTCs requires that $f_1>0$, and so a necessary
condition for no naked CTCs is that on the horizon, the expression
on the second line be non-negative.   This can be satisfied if 
$s^2<s^2_0\equiv 1/(\Xi_+^2-1)$,  provided that
$m$ is sufficiently large.  However, in the BPS limit, where $s=s_0$, 
the metric will necessarily have naked CTCs.  (In fact recently an
alternative supersymmetric limit of our seven-dimensional
non-extremal black hole solution has been found,
which does include a regular black hole
with no CTCs or singularities on or outside the event horizon \cite{cglp}.)

\section{Conclusions}

    In this paper, we have constructed non-extremal charged rotating
black hole solutions in seven-dimensional gauged supergravity.  The
solutions carry two independent charges, associated with gauge fields
in the $U(1)\times U(1)$ abelian subgroup of the $SO(5)$ gauge
group. In order to simplify the problem we set the three angular
momenta of a generic rotating black hole equal. An interesting new
feature that arises in seven dimensions is that the 4-form field
$F_\4$, which is also non-zero when the two electric charges are
both non-vanishing, satisfies a first-order ``odd-dimensional self-duality''
equation.  This implies that the structure of the solutions is considerably
more complicated than in previous examples that were studied in four and
in five dimensions.   As well as obtaining the non-extremal black hole
solutions, we also considered their BPS limits, showing that one
can obtain supersymmetric rotating black hole solutions of
seven-dimensional gauged supergravity.

   The results presented in this paper are of significance for the
AdS$_7$/CFT$_6$ correspondence in M-theory.

\section*{Acknowledgments} M.C. is grateful to the George P. \& 
Cynthia W. Mitchell Institute for Fundamental Physics for hospitality during
the course of this work.

\end{document}